\documentclass[preprint]{emulateapj}

\slugcomment{Accepted for publication in the Astrophysical Journal Letters}
\shortauthors{Zaritsky, Gonzalez, \& Zabludoff}
\shorttitle{Dwarfs and the Fundamental Manifold}

\begin{document}
\title{
Local Group Dwarf Galaxies and the Fundamental Manifold of Spheroids}
  
\author{Dennis Zaritsky\altaffilmark{1}, Anthony H. Gonzalez\altaffilmark{2}, and Ann I. Zabludoff\altaffilmark{1} }
\altaffiltext{1}{Steward Observatory, University of Arizona, 933 North Cherry Avenue, Tucson, AZ 85721}
\altaffiltext{2}{Department of Astronomy, University of Florida, Gainesville, FL 32611}


\begin{abstract}    

The fundamental manifold (FM), an extension of the fundamental
plane formalism, incorporates all spheroid-dominated stellar systems from dwarf ellipticals up to the 
intracluster stellar populations of  galaxy clusters by accounting for the 
continuous variation of the mass-to-light ratio within the effective
radius $r_e$ with scale. 
Here, we find that Local Group dwarf spheroidal and dwarf elliptical galaxies,
which probe the FM relationship roughly one decade lower in $r_e$ than previous work,
lie on the extrapolation of the FM.  When combined with the earlier data, 
these Local Group dwarfs demonstrate the validity of the
empirical manifold over nearly four orders of magnitude in $r_e$.
The continuity of the galaxy locus on the manifold and, more specifically, the overlap 
on the FM of dwarf ellipticals like M 32 and dwarf spheroidals like Leo II, implies that
dwarf spheroidals belong
to the same family of spheroids as their more massive counterparts.
The only significant outliers are
Ursa Minor and Draco. We explore whether the deviation of these two galaxies from the 
manifold reflects a breakdown in the coherence of the 
empirical relationship at low luminosities or rather 
the individual dynamical peculiarities of these
two objects. 
We discuss some implications of our results for how the lowest
mass galaxies form.
\end{abstract}


\section{Introduction}

Scaling relationships among galaxies provide many of the principal constraints on galaxy
evolution models. As such, the ``fundamental manifold" of spheroids \citep[hereafter FM;][]{zgz}, which spans
a factor of 1000 in effective radius $r_e$ and of 100 in velocity dispersion $\sigma$, poses 
a challenge to models of the formation of kinematically hot stellar components
embedded in dark matter halos.
\cite{zgz} focused principally on relating the structural
properties of the spheroidal stellar component of the largest known virialized systems, the intracluster stars of 
galaxy clusters \citep[CSph;][]{gzz}, to those of more common spheroids such as giant (E) and dwarf
(dE) elliptical galaxies. They 
considered the efficiency with which baryons are packed relative to dark matter within $r_e$
(as measured by the mass-to-light ratio within $r_e,  M_e/L_e$) and found that the existence
of the spheroid FM implies a $M_e/L_e - \sigma$ relationship that continues to steepen
with $\sigma$ from Es to CSphs. In other words, the packing of baryons relative to dark
matter within $r_e$ is less efficient for these spheroids as $\sigma$ increases. The behavior for 
spheroids with still smaller values of $\sigma$, dEs and dwarf Spheroidals (dSphs),
was inconclusive.
In this Letter, we consider this low-$\sigma$ extreme of spheroid scale, using eight
Galactic dSphs and four dE companions of M 31 to investigate
how well the smallest and most compact galaxies map onto the extrapolation of the FM.
Finally, we speculate on how the behavior of the FM for these lowest mass galaxies
might be used to constrain galaxy formation models.

\label{sec:intro}

\section{Results and Discussion}

\begin{figure}
\plotone{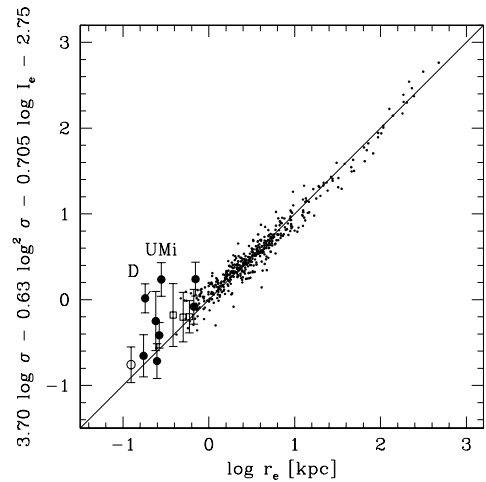}
\caption{Local Group spheroids plotted on the fundamental manifold. The small dots represent the data
presented in \cite{zgz}, which include the intracluster stellar component of galaxy clusters (CSph), brightest cluster galaxies (BCG), giant
ellipticals (E), and dwarf ellipticals (dE). The filled circles with errorbars represent the eight Galactic dwarf
spheroidals, with the two vertically outlying points representing Ursa Minor (UMi) and Draco (D). 
The open symbols
represent the four M 31 satellites (the open circle at the bottom left of the Figure represents M 32).
The solid line is the edge-on projection of the manifold in this
coordinate system as derived by \cite{zgz}. The short diagonal line extending from the Draco data point
represents the change in position caused by adopting the structural parameters of \cite{ode} rather
than those from our calculations using the \cite{ih} King model fit.
}
\label{fig:fm}
\end{figure}

We searched the literature for the necessary data for every dwarf
elliptical and spheroidal in the Local Group. 
The data (Table 1, Figure \ref{fig:fm}) come from various sources.
We calculate 
the half light radii for the Galactic dwarf spheroidals using the King model
fit parameters given by \cite{ih}. For the Andromeda systems, we adopt the $r_e$ 
given by \cite{kent} for M 32 and NGC 147, fit to Kent's photometry for
NGC 185, and use the value given by \cite{choi} for NGC 205. The mean surface
brightness within $r_e$, $I_e$, is evaluated using the total luminosity and a circular
aperture defined by $r_e$.
For consistency with the data used by \cite{zgz}, we convert magnitudes 
from $V$ to Cousins $I$ using the
color transformation for ellipticals from \cite{fukugita}. We do not attempt to correct $r_e$
for any possible color dependencies.
The errorbars represent the propagated uncertainty due to the quoted statistical errors in
 $\sigma$, the apparent magnitude, and the distance.  Errors in
the half-light or effective radius are not included as they affect both the ordinate
and abscissa such that the data points move nearly parallel to the projection
of the fundamental manifold (see Figure 1 for a demonstration). The velocity dispersion uncertainties are the dominant source of the plotted errors.

We compare the position of these 12 dwarfs to the FM
of \cite{zgz} in Figure 1. We have not modified the old relationship to optimize the fit to 
this new set of points. 
The $\chi^2_\nu$ value of only these new data with respect to the existing model is 1.91. This value
is high, even for the relatively small sample given here (such a value can be ruled
out as random with $>$ 98\% confidence). However, this apparent discrepancy is due solely to 
Ursa Minor (UMi) and Draco (without these two galaxies, 
$\chi^2_\nu$ drops to 0.82).

Apart from the two outliers, which are discussed in more detail below,
the other 10 systems extend the FM relation
nearly a decade lower in effective radius.
The continuity of the locus of spheroids on the FM, from CSphs to dEs,  
suggested a single family of spheroids \citep{zgz}.
The overlap here between dEs and the 
Galactic dSphs indicates that even dSphs are part of this family. A further example of
the continuity of spheroid properties on the FM is
that one of the highest surface brightness galaxies known, M 32, lies next to
one of the lowest surface brightness galaxies, Leo II. These two galaxies also differ
in velocity dispersion by nearly a factor of ten. M 32's presence on the FM 
suggests that unless tidal stripping works to move objects along the FM, M 32 has
been only modestly disturbed by its interaction with M 31 
\citep[a similar conclusion is reached through other arguments by][]{choi}.  

We note that even for the 10 dwarfs that are statistically consistent with
the FM extrapolation, there may be a small but
systematic deviation. All but one of the dwarf
galaxies lie above the FM projection, while at larger values of $r_e$ ($1 < \log r_e  < 2$, $r_e$ in kpc) the
data appear to lie systematically below the line. These systematic deviations may indicate
that there is a more precise 
fit to the global FM than that presented by \cite{zgz}. However, given the uncertainties in the measurements (see
the scatter in published $m_V$ values for the Galactic dSphs in Table 1), the possible systematic
behavior of $r_e$ with color, the use of a single universal color term across the entire
range of systems, and the difficulties in defining $\sigma$, it is premature
to fine-tune the FM coefficients.

Returning to the two outliers, Draco and UMi, we discuss
three possible causes (in order of least to most astronomically interesting):
1) the difficulty in measuring the basic fundamental parameters of these systems has been 
underestimated, leading to overly optimistic errorbars or to systematic errors,
2) these two galaxies are experiencing tidal forces that affect their internal kinematics
and structure such that they no longer lie on the FM, or
3) these two galaxies, which have the lowest luminosities of all the galaxies
in the sample, mark the breakdown of the tight FM relationship at low luminosities,
and so identify the location in this parameter space at which baryons and dark matter no longer
follow the same ``rules" of galaxy formation.

\begin{deluxetable*}{lrrrrrrr}
\tabletypesize{\scriptsize}
\tablecaption{Local Group Spheroids}

\label{tab:data}
\tablewidth{0pt}
\tablehead{
\colhead{Name} &
\colhead{$r_e$}   &
\colhead{$m_V$}   &
\colhead{$m_V$ Range}   &
\colhead{$\sigma_v$}   &
\colhead{$D$}   &
\colhead{References} \\
&
\colhead{[arcmin]}   &
   &
   &
\colhead{[km s$^{-1}$]} &
\colhead{[kpc]}  \\
}

\startdata

Ursa Minor&14.5$\pm$1.1&10.3$\pm$0.4&9.8-10.69&8.8$\pm$0.8&66$\pm$3&1, 9, 12\\
Draco&8.2$\pm$0.7&10.9$\pm$0.3&10.78-11.0&8.5$\pm$0.7&76$\pm$6&1, 5, 9, 15\\
Sculptor&11.5$\pm$1.9&8.5$\pm0.3$&8.13-8.8&8.8$\pm$0.6&79$\pm$4& 9, 12, 19, 20\\
Sextans&28.0$\pm$5.8&10.3$\pm$0.3&10.2-10.4&6.2$\pm$0.7&86$\pm$4&9, 12, 16\\
Carina&8.2$\pm$1.1&10.85$\pm$0.25&10.6 -11.0&6.8$\pm$1.6&101$\pm$5&9, 11, 12, 13\\
Fornax&16.7$\pm$1.0&7.6$\pm$0.3&6.9-8.4&12.4$\pm$1.5&138$\pm$8&9, 12, 19\\Leo II&2.6$\pm$0.4&12.0$\pm0.2$&11.6-12.0&6.7$\pm$1.1&233$\pm$12&3, 9, 12, 18\\
Leo I&3.4$\pm$0.4&10.1$\pm$0.3&10.0-10.2&8.8$\pm$0.9&254$\pm$30&2, 9, 12, 14\\
NGC 185&2.50&9.09$\pm$0.15&...&28$\pm$8&620$\pm$25&8, 10, 12\\
NGC 147 &2.13&9.35$\pm$0.15&...&23$\pm$5&725$\pm$45&4, 10, 12\\
M 32 &0.53&8.10$\pm$0.15&...&60$\pm$10&805$\pm$35&6, 10, 12, 17\\
NGC 205 &2.38&8.05$\pm$0.15&...&35$\pm$5&815$\pm$35&6, 7, 10, 12\\

\enddata

\tablecomments{The effective radii were determined
using the King model fit parameters from \cite{ih} for the Galactic dSphs and from the data
of \cite{kent} for NGC 185. Other data are drawn directly from the references.}

\tablerefs {[1] \cite{arm95},  [2] \cite{bellazini04},  [3] \cite{bellazini05},  [4] \cite{bender}, [5] \cite{bonanos},
 [6] \cite{choi},  [7] \cite{geha},   [8] \cite{held},  [9] \cite{ih}, [10] \cite{kent},  [11] \cite{mateo93}, [12] \cite{mateo98}, [13] \cite{mateo98b}, [14] \cite{mateo98a}, [15] \cite{ode}, [16] \cite{suntzeff},  [17] \cite{vdm2}, [18] \cite{vogt}, [19] \cite{walker}, [20] \cite{westfall}}

\end{deluxetable*}

If the offset between these two galaxies and the FM is due to Option 1, 
then the large value of $\chi^2_\nu$ 
must be attributed to a problem with the 
velocity dispersion measurements and/or their estimated uncertainties.
Errors in the apparent magnitude and distance are unlikely
to cause such a large discrepancy. For example, to bring 
Draco back onto the FM
requires an error in the absolute magnitude of 2.86 mag. Even though one could parcel
this error into errors in the apparent magnitude, color conversion, and distance, these errors
would still be unreasonably large.
Alternatively, one could place Draco on the FM with a modest change of 
its velocity dispersion from 8.5 \citep{arm95} to 5 km sec$^{-1}$.  Is 
such a systematic error likely?

Measurements of the velocity dispersions of dSphs have improved greatly
over the past two decades and now come from samples of hundreds of 
stars. The quoted uncertainty on the measurement we use is 0.7 km sec$^{-1}$ \citep{arm95}, so
the required change in $\sigma$ from 8.5 to 5 km s$^{-1}$ appears unlikely on purely statistical grounds. 
Recent, larger studies \citep{kleyna02, munoz} confirm 
the  global velocity dispersion from \cite{arm95}, although they do find radial gradients in $\sigma$.
The velocity dispersion used in the FM relationship is supposed
to be that measured within $r_e$, although we generally assume
flat dispersion profiles and adopt the
measured global value. The dispersion profile presented by \cite{munoz} suggests that
$\sigma$ could be as low as $\sim$ 6 km sec$^{-1}$ over a centrally limited
radius, which could help to reconcile Draco's position relative to the FM. However,
UMi has a velocity dispersion gradient of the opposite sign, so its central $\sigma$ is larger than the global
value. Taking a more central value for $\sigma$ would increase
the discrepancy between UMi's location in Figure 1 and the FM.
We conclude that although the newly identified velocity dispersion gradients complicate the
picture (and so perhaps lessen the discrepancy by increasing the errorbars), they do not
act in a systematic manner to address the current discrepancy between the FM and the properties 
of UMi and Draco.

Proceeding to Option 2, we can speculate that these two systems are not well described by 
simple dynamics.
They are the two closest galaxies to the Milky Way among our set, as well as the two of 
lowest luminosity.
Because of these properties, it is often suggested that 
Draco and UMi are victims of tidal
stripping \citep{km, kuhn, kroupa, fleck, gomez, munoz}. For example, \cite{munoz} 
argue that this galaxy is not a simple, dynamically relaxed system because of the 
unphysically large inferred mass (implying $M/L >$ 900) based on their measurement of a large
$\sigma$ far from UMi's core. Such
arguments favor the interpretation that the dynamics of these systems are somewhat disturbed,
that the velocity dispersions are affected, and that these systems should
lie off the FM. The critical test of this interpretation is to determine where other 
low luminosity spheroids, especially those 
far from the Milky Way or M31 and thus unperturbed by tidal forces, fall with respect to the FM.

There are at least two other Local Group objects that are particularly interesting for 
this discussion given their low luminosity: the Ursa Major dwarf \citep{willman05}
and And IX \citep{zucker}. Unfortunately,
the data are not yet of the quality necessary to impact this discussion. The magnitude
and structural parameters of UMa are currently crude estimates, although improved
data are on the way. If we do adopt the published structural parameters \citep{willman05, kleyna05}, then UMa lies well above the
FM, as do UMi and Draco. The data for And IX are further along \citep{harbeck,chapman}, but the velocity dispersion
is still somewhat problematic. The stellar velocity dispersion is almost zero at small radii and
then rises quickly with radius.
The global velocity dispersion value may therefore be misleading. Nevertheless, if we
adopt the published value, we find that And IX is also above the FM relationship, although 
only by 1.1 standard deviations given the large uncertainty. Improving the data on these two galaxies
is key because they probe the low-luminosity end of the spheroid distribution. 

Finally, considering Option 3, we restate that 
Draco and UMi are the lowest luminosity galaxies plotted in Figure \ref{fig:fm} and that the
preliminary data discussed above for two other low luminosity dwarfs show consistent deviations from the FM. As such, 
Draco and UMi might be indicative of a breakdown in the coherence of the 
FM at these luminosities.  Do we expect such a breakdown?

As discussed by \cite{zgz}, the trend of $M_e/L_e$ with $\sigma$ that leads
to the FM can be physically interpreted as a trend in 
the packing efficiency of baryons relative to dark matter within $r_e$. They concluded
that the packing is inefficient for the systems with largest $\sigma$ (CSphs and BCGs), becomes 
highly efficient for systems with moderate $\sigma$ (Es), and, with less certainty, becomes inefficient once again in systems with low $\sigma$ (dEs). The continuation of the FM relationship
here to even lower $\sigma$ systems confirms that the packing of baryons relative
to dark matter is highly inefficient in the systems with the lowest $\sigma$'s.
The key to a tight FM
lies in the maintaining this pattern in the relative distributions
of baryons and dark matter among all spheroids. If there is a class
of spheroids for which a change in the distribution of baryons due to non-gravitational
physics is not accompanied by 
a corresponding change in the distribution of the dark matter,  we expect a loss of the FM coherence for those spheroids. 

One such class may be 
systems in which the baryons are dynamically negligible and that therefore
lack a physical connection between
the properties of the baryonic
($r_e$, $I_e$) and dark ($\sigma$) components. If the baryons in these
systems are subject to forces that do not affect the dark matter, such as winds or ram pressure
stripping, the dark matter would not respond to changes in baryon distribution.
In other words, the baryons can be distributed, or packed,
within the dark matter halo in numerous ways without affecting $\sigma$, 
and this range of optical $r_e$'s and $I_e$'s will almost certainly break the coherence of the FM.

The baryons are dynamically negligible in low-luminosity systems
with large mass-to-light ratios 
\citep[see][]{mateo98} and where non-gravitational physics is expected to be important \citep{br,martin}. 
As one example of how the FM coherence may be broken,
we consider a population of dark-matter-dominated, low-luminosity systems in which
varying fractions of baryons have been lost due to supernova winds. These systems will all have the
same $\sigma$ (because the dark matter properties are unaffected by the baryon loss), 
but different values of $I_e$. Unless $r_e$ conspires to 
compensate for the change in $I_e$, this set of objects will lie off the FM by differing
amounts.
Upward scatter in
Figure \ref{fig:fm}, as seen for Draco and UMi, corresponds to a lower than predicted
effective surface brightness, as would arise if these dSphs are dark-matter-dominated
and have experienced baryon loss.

The adherence or deviation of low-mass galaxies from the FM may 
ultimately help to address  
one of the key questions facing current hierarchical
models --- how do galaxies populate low mass halos? A wide range of literature
has considered the ``missing satellite problem" formalized by \cite{moore} in 
which cold dark matter cosmological models apparently overpredict the number of 
low luminosity satellite galaxies.
It is now evident that any solution to this problem must include
a delicate balancing act
to retain the FM well into the low mass regime where the problem has been identified.
The tightness of the FM relationship down to at least the luminosity of UMi and Draco
suggests that solving the missing satellite problem may require investigating two regimes,
namely those in which baryons are and are not dynamically important.
Processes that inherently generate significant scatter, such as any that are environmentally
driven \citep[see, for example, ][]{kravtsov}, could easily fail to reproduce the low scatter seen in the 
FM for systems more luminous than UMi and Draco.
If the other hand, we were able to quantify the scatter along the vertical axis in Figure \ref{fig:fm}
for objects like UMi and
Draco, then that scatter would constrain the variations in mass loss or 
star formation efficiency
allowed as a function of $\sigma$.

\section{Summary}

We demonstrate that the lowest luminosity, lowest surface brightness, and lowest velocity
dispersion spheroidal galaxies currently known fall on the projection of the ``fundamental manifold" (FM) defined
by \cite{zgz}. The FM now spans nearly four orders of magnitude in effective radius $r_e$.
The FM is not a simple 
recasting of the virial theorem, as demonstrated by the
complex behavior of the mass-to-light ratio within $r_e$ with $\sigma$  \citep{zgz} that is now extended
by these new data.
The FM
 places a constraint on models of spheroid formation ranging from the dwarf spheroidals
of the Local Group to the intracluster stellar component of rich galaxy clusters. In particular,
the manifold describes an increase in 
the ratio of dark to luminous matter within the optically defined effective radius
for both the largest and smallest spheroids embedded in dark matter halos. 
The continuity of this relationship into the 
regime of dSphs suggests that these systems are not a distinct class of 
spheroid and that they have not been grossly affected by interactions with their massive
parent galaxy. The tightness of the FM well into the low mass regime, where the ``missing
satellite" problem arises \citep{moore}, suggests that it may be challenging to explain
the existing systems as the few sole survivors of a complex and violent history.

\begin{acknowledgments}
The authors thank Ed Olszewski for advice regarding the existing literature on dSphs.
DZ acknowledges financial support for this work from the David and Lucile Packard Foundation,
NASA LTSA award NNG05GE82G and NSF grant AST-0307482.
AIZ acknowledges financial support from NASA LTSA award NAG5-11108 and
from NSF grant AST-0206084.
\end{acknowledgments}


\begin{thebibliography}{}

\bibitem[Armandroff et al.(1995)]{arm95}
Armandroff, T.~E., Olszewski, E.~W. \& Pryor, C. 1995, \aj, 110, 2131

\bibitem[Babul \& Rees(1992)]{br}
Babul, A., \& Rees, M.~J. 1992, \mnras, 255 346

\bibitem[Bellazzini et al.(2005)]{bellazini05}
Bellazzini, M.,  Gennari, N.,  \& Ferraro, F.~R. 2005, \mnras, 360, 185

\bibitem[Bellazzini et al.(2004)]{bellazini04}
Bellazzini, M.,  Gennari, N.,  Ferraro, F.~R., \& Sollima, A. 2004, \mnras, 354, 708

\bibitem[Bender et al.(1991)]{bender}
Bender, R.,  Paquet, A.,  \& Nieto, J.-L.1991 \aap, 246, 349

\bibitem[Bonanos et al.(2004)]{bonanos}
Bonanos, A.~Z.,  Stanek, K.~Z.,  Szentgyorgyi, A.~H., Sasselov, D.~D., \&  Bakos, G.~A. 2004, \aj, 127, 861

\bibitem[Chapman et al.(2005)]{chapman}
Chapman, S.~C., Ibata, R.,  Lewis, G.~F.,  Ferguson, A.~M.~N.,
Irwin, M.,  McConnachie, A., \&  Tanvir, N. 2005, \apj, 632, 87L

\bibitem[Choi et al.(2002)]{choi}
Choi, P.~I.,  Guhathakurta, P., \&  Johnston, K.~V. 2002, \aj, 124, 310


\bibitem[Fleck \& Kuhn(2003)]{fleck}
Fleck, J.-J., \& Kuhn, J.~R. 2003, \apj, 592, 147

\bibitem[Fukugita et al.(1995)]{fukugita}
Fukugita, M., Shimasaku, K., \& Ichikawa, T. 1995, \pasp, 107, 945

\bibitem[Geha et al.(2006)]{geha}
Geha, M.,  Guhathakurta, P.,  Rich, R.~M., \&  Cooper, M.~C. 2006, \aj, 131, 332

\bibitem[G{\'o}mez-Flechoso \& Mart{\'{\i}}nez-Delgado(2003)]{gomez}
{G{\'o}mez-Flechoso}, M.~{\'A}. \& {Mart{\'{\i}}nez-Delgado}, D. 2003, \apj, 586, 123L


\bibitem[Gonzalez et al.(2005)]{gzz}
Gonzalez, A.~H., Zabludoff, A.~I., \& Zaritsky, D. 2005, \apj,  618, 195

\bibitem[Harbeck et al.(2005)]{harbeck}
Harbeck, D., Gallagher, J.~S., Grebel, E.~K., Koch, A. , \& Zucker, D.~B. 2005, \apj, 623, 159

\bibitem[Held et al.(1992)]{held}
Held, E.~V.,  de Zeeuw, T.,  Mould, J., \&  Picard, A. 1992, \aj, 103, 851

\bibitem[Irwin \& Hatzidimitriou(1995)]{ih}
Irwin, M. \& Hatzidimitriou, D.,  1995, \mnras, 277, 1354

\bibitem[Kent(1987)]{kent}
Kent, S.~M., 1987, \aj, 94, 306


\bibitem[Kleyna et al.(2005)]{kleyna05}
Kleyna, J.~T., Wilkinson, M.~I., Evans, N.~W., \& Gilmore, G. 2005, \apj, 630, 141L

\bibitem[Kleyna et al.(2002)]{kleyna02}
Kleyna, J.,  Wilkinson, M.~I.,  Evans, N.~W.,  Gilmore, G.,
\& Frayn, C. 2002, \mnras, 330, 792

\bibitem[Kravtsov et al.(2004)]{kravtsov}
Kravtsov, A.~V., Gnedin, O.~Y., \& Klypin, A.~A. 2004, \apj, 609, 482

\bibitem[Kroupa(1997)]{kroupa}
Kroupa, P. 1997, New Astronomy, 2, 139

\bibitem[Kuhn(1993)]{kuhn}
Kuhn, J.~R. 1993, \apj, 409, 13L

\bibitem[Kuhn \& Miller(1989)]{km}
Kuhn, J.~R., \& Miller, R.~H. 1989, \apj, 341, 41L

\bibitem[Martin(1999)]{martin}
Martin, C.~L. 1999, \apj, 513, 156


\bibitem[Mateo(1998)]{mateo98}
Mateo, M. 1998, Ann. Rev. 36, 435

\bibitem[Mateo et al.(1998b)]{mateo98b}
Mateo, M.,  Hurley-Keller, D., \&  Nemec, J. 1998, \aj, 115, 1856

\bibitem[Mateo et al.(1993)]{mateo93}
Mateo, M.,  Olszewski, E.~W.,  Pryor, C.,  Welch, D.~L., \& Fischer, P. 1993, \aj, 105, 510

\bibitem[Mateo et al.(1998a)]{mateo98a}
Mateo, M., Olszewski, E.~W., Vogt, S.~S., \& Keane, M.~J. 1998, \aj, 116, 2315

\bibitem[Moore et al.(1999)]{moore}
Moore, B., Gingna, S., Governato, F., Lake, G., Quinn, T., Stadel, J., and Tozzi, P. 1999, \apj, 524, L19

\bibitem[Mu{\~n}oz et al.(2005)]{munoz}
Mu{\~n}oz, R.~R. et al. 2005, \apj, 631, 137L


\bibitem[Odenkirchen et al.(2001)]{ode}
Odenkirchen, M. et al. 2001, \aj, 122, 2538


\bibitem[Suntzeff et al.(1993)]{suntzeff}
Suntzeff, N.~B., Mateo, M.,  Terndrup, D.~M.,  
	Olszewski, E.~W.,  Geisler, D., \&  Weller, W. 1993, \apj, 418, 208


\bibitem[van der Marel et al.(1994b)]{vdm2}
van der Marel, R.~P.,  Evans, N.~W.,  Rix, H.-W.,  
White, S.~D.~M.,  \& de Zeeuw, T. 1994, \mnras, 271, 99


\bibitem[Vogt et al.(1995)]{vogt}
Vogt, S.~S., Mateo, M., Olszewski, E.~W., \&  Keane, M.~J. 1995, \aj, 109, 151

\bibitem[Walker at al.(2006)]{walker}
Walker, M.~G., Mateo, M., Olszewski, E.~W., Bernstein, R.~A., Wang, X., \& Woodroofe, M. 2006, 
\aj, in press (astro-ph/0511465)

\bibitem[Westfall et al.(2006)]{westfall}
Westfall, K.~B., Majewski, S.~R., Ostheimer, J.~C.,
	Frinchaboy, P.~M., Kunkel, W.~E., Patterson, R.~J., \&
	Link, R. 2006, \aj, 131, 375


\bibitem[Willman et al.(2005)]{willman05}
Willman, B., et al. 2005, \apj, 626, 85L

\bibitem[Zaritsky et al.(2006)]{zgz}
Zaritsky, D., Gonzalez, A.~H., \& Zabludoff, A.~I. 2006, \apj, 638, 725

\bibitem[Zucker et al.(2004)]{zucker}
Zucker, D.~B., et al. 2004, \apj, 612, 121L

\end{thebibliography}
\end{document}